\documentclass[aps,pra,floatfix,twocolumn,superscriptaddress]{revtex4-1}
\usepackage[final]{graphicx}
\usepackage{times,amsmath,amssymb}
\usepackage{epsfig,color}
\usepackage{hyperref}
\usepackage[caption = false]{subfig}
\usepackage{thumbpdf,enumerate}
\usepackage{booktabs}

\newcommand{\Tr}{{\rm Tr}}

\def\Tr{\hbox{Tr}} 
\usepackage{pifont}

\newcommand{\ket}[1]{\vert#1\rangle}
\newcommand{\bra}[1]{\langle#1\vert}

\begin{document}

\title{Realism-information complementarity in photonic weak measurements}

\author{Luca Mancino}
\affiliation{Dipartimento di Scienze, Universit\`{a} degli Studi Roma Tre, Via della Vasca Navale 84, 00146, Rome, Italy}

\author{Marco Sbroscia}
\affiliation{Dipartimento di Scienze, Universit\`{a}  degli Studi Roma Tre, Via della Vasca Navale 84, 00146, Rome, Italy} 

\author{Emanuele Roccia}
\affiliation{Dipartimento di Scienze, Universit\`{a} degli Studi Roma Tre, Via della Vasca Navale 84, 00146, Rome, Italy}

\author{Ilaria Gianani}
\affiliation{Dipartimento di Scienze, Universit\`{a} degli Studi Roma Tre, Via della Vasca Navale 84, 00146, Rome, Italy}

\author{Valeria Cimini}
\affiliation{Dipartimento di Scienze, Universit\`{a} degli Studi Roma Tre, Via della Vasca Navale 84, 00146, Rome, Italy}

\author{Mauro Paternostro}
\affiliation{Centre for Theoretical Atomic, Molecular and Optical Physics, School of Mathematics and Physics, Queen's University Belfast, Belfast BT7 1NN, United Kingdom}

\author{Marco Barbieri}
\affiliation{Dipartimento di Scienze, Universit\`{a} degli Studi Roma Tre, Via della Vasca Navale 84, 00146, Rome, Italy}
\affiliation{Istituto Nazionale di Ottica - CNR, Largo Enrico Fermi 6, 50125, Florence, Italy}

\begin{abstract} 
The emergence of realistic properties is a key problem in understanding the quantum-to-classical transition. In this respect, measurements represent a way to interface quantum systems with the macroscopic world: these can be driven in the weak regime, where a reduced back-action can be imparted by choosing meter states able to extract different amounts of information. Here we explore the implications of such weak measurement for the variation of realistic properties of two-level quantum systems pre- and post-measurement, and extend our investigations to the case of open systems implementing the measurements.
\end{abstract}

\maketitle

\section{Introduction}  
The classical interpretation of the result of a measurement is merely the disclosure of a property of the system at the moment of its observation. It is now clear that this view fails to capture the more intricate process of measuring a quantum object. The latter breaks the normal evolution of the quantum state of the object, and results in the observable assuming instantaneously the measured value. An extensive body of literature has been dedicated to discussing this matter, with interpretations ranging from the standard operative Copenhagen view (``shut up and calculate"~\cite{Merm89}), to Bohmian mechanics~\cite{Bohm52}, to the Bayesian concept of the state collapse as an information update~\cite{Cave02}, to more exotic suggestions such as the many-world theory~\cite{Ever57} and collapse models~\cite{GRW,Bass13}. Regardless of the preference to the possible solution of the measurement problem, we are confronted with the need to understand how the classical world, where realistic values for observable might not be inherent, but are certainly tenable, emerges from the quantum un-realistic world.
 
In this debate, a prominent role is reserved to the notion of elements of the reality, that Einstein, Podolsky and Rosen introduced as intrinsic properties of the system that can be predicted with certainty without any disturbance~\cite{EPR35}. This notion complemented with that of locality is unable to explain peculiar quantum phenomena such as entanglement~\cite{Bell64,Geno05}. These elements are generally associated to the wavefunction, the only description of the reality quantum mechanics is able to provide. The current debate is centred on whether the wavefunction itself has an ontic nature, {\it i.e.} it has a realistic connotation, or it is merely epistemic paraphernalia to describe an underlying realistic nature~\cite{Puse12, Colb12, Lewi12,Patr13,Leif13,Barr14,Branc14,Ring15,Nigg12}.

Recently, Bilobran and Angelo introduced a notion of realism based on both quantum states and observables, and connected it to an experimental procedure~\cite{Bilo15}: an element of the reality is introduced for the observable $O$ whenever the quantum states, here considered as a density matrix $\rho$, is not altered by a measurement of $O$; this adheres to the standard notion of classicality as that of a state that commutes with any measurement operator. A measurement of the realistic content of $\rho$ is then defined based on the entropy of the pre- and post-measurement states. If weak monitoring replaces the projective measurement~\cite{Ahar90}, it has been shown that under ideal conditions the change in the entropic measure of reality content of $\rho$, $\Delta{ \mathcal R}$, is in a duality relation with the amount of information extracted, $\Delta I$~\cite{Dieg18}:
\begin{equation}
\Delta{ \mathcal R}+\Delta I=0,
\label{BA}
\end{equation}
close to those introduced in Ref.~\cite{Baga16} for coherence and which-path. In this article, we explore this relation in an experiment based on a photonic weak measurement device~\cite{Pryd04,Pryd05,Scia06,Barb09,Kim12,Roze12,Piac16,Kim18,Manc18}. We show to what extent this equality can guide observation in actual experiments. In addition, we extend our investigations to the case of an open-system implementing a quantum measurement~\cite{Manc18}, and draw considerations on how the initial entropy connected to the measuring device connects to the emergence of realistic characters, according to the definition of Bilobran and Angelo. 

\section{A measure of reality.} Our intent is to investigate how a realistic description becomes possible as we tune the invasivity of the measurement from negligible (weak measurement) to the standard projective regime (strong measurement)~\cite{Ahar90}. Therefore, the figure of merit we must use should be capable of addressing mixtures and should be related to measurable quantities. We consider the case where an observable $O$ is measured on a generic quantum state $\rho$ by means of a suitable device. The definition in Ref.\cite{Bilo15} considers the degree of irreality of the observable $O$, described in quantum mechanics by the operator $\hat O = \sum_k o_k \ket{k}\bra{k}$, associated to the state $\rho$ as: 
\begin{equation}
{\mathcal I}(O|\rho)=S(\Phi_{O}(\rho))-S(\rho),
\label{irreality}
\end{equation}
where $S$ is the Von Neumann entropy and the map $\Phi_{O}(\rho)=\sum_k p_k \ket{k}\bra{k}$, with $p_k=\bra{k}\rho\ket{k}$, describes the action of the measuring device. The degree of irreality of $O$ vanishes if this can be measured without affecting the state, and it is maximum when the measurement of $O$ is disruptive to the point of bringing a pure state into a complete mixture; the latter correspond to a case in which $\rho$ is an equal superposition of all possible eigenstates $\ket{k}$. This definition then reveals an epistemic approach, as it is only concerned with our ignorance of the realistic value of the observable $O$.

We need to extend these positions to the case of a weak measurement: this is a generalisation of the standard projective measurement, for which the output state is not unambiguously identified, although some information is obtained. The implementation of a weak measurement is typically carried out by coupling the system with a pointer object, which is then measured~\cite{Ahar90}. Due to the coupling, the value of the observable $O$ modifies the distribution of a related observable $Q$ on the pointer. When the effect of such modification allows to discriminate different states of the pointer - {\i.e.} the induced shift of the mean value of $Q$ is significantly larger than the width $\delta Q$ - the measurement functions in the standard conditions. In opposite limit in which the size of the shift is comparable to $\delta Q$, we operate in the weak regime. For any measurement strength, an element of the reality can be defined whenever there exist a procedure to predict with certainty what shift will occur: the element can then attributed to the shift itself~\cite{Vaid96}.

 The generalisation of the map $\Phi_O$ to the weak regime is performed as follows~\cite{Bilo15,Dieg18}. Upon collecting the outcome $k$, the state emerging from the weak measurement is written as: ${\mathcal C}^\epsilon_k(\rho) = (1-\epsilon)\rho+\epsilon \ket{k}\bra{k}$. The limit $\epsilon{\rightarrow}1$ corresponds to the projective case extracting maximal information, and $\epsilon{\rightarrow}0$ corresponds to performing to measurement that clearly delivers no information. If the outcomes are not sorted, the average state is ${\mathcal M}^\epsilon_O(\rho)=\sum_k p_k {\mathcal C}^\epsilon_k(\rho) $~\cite{Dieg18}. The map ${\mathcal M}^\epsilon_O$ has the remarkable property of commuting with $\Phi_O$: ${\mathcal M}^\epsilon_O \Phi_O = \Phi_O {\mathcal M}^\epsilon_O$ for all strengths. This implies that the map ${\cal M}^\epsilon_O(\rho)$ can not be invoked as introducing any element of the reality, whenever $\Phi_O$ did not.

We can use these definitions to calculate the variation of the degree of reality of $O$ following a weak measurement as~\cite{Bilo15}:
\begin{equation}
\begin{aligned}
\Delta{\mathcal R} =& -\Delta{\mathcal I} = I(O|\rho)-I(O|{\mathcal M}^\epsilon_O(\rho))\\
=&\, S({\mathcal M}^\epsilon_O(\rho))-S(\rho),
\label{deltaR}
\end{aligned}
\end{equation}
where we have used the definition of irreality \eqref{irreality} and the commutation properties of ${\mathcal M}^\epsilon_O$ to obtain the last equality. By invoking the concavity of the Von Neumann entropy, the variation Eq.~\eqref{deltaR} can be bound as~\cite{Dieg18}:
\begin{equation}
\Delta{\mathcal R}\geq \epsilon\,{\mathcal I}(O|\rho),
\label{bound}
\end{equation}
demonstrating that the degree of reality of $O$ is always non decreasing upon monitoring.

\section{Reality-information duality} 

The variation of the degree of reality can be directly related to a change in the information content of the initial state $\rho$~\cite{Dieg18}. In order to define a proper quantifier, we analyse the measurement strategy in detail. The weak monitoring is performed by introducing an ancillary system $\rho_A =\ket{A}\bra{A}$, and coupling it to the system by means of the unitary dynamics $\hat U$:
\begin{equation}
{\mathcal M}^\epsilon_k(\rho) = \Tr_A\left( \hat U \rho\otimes\ket{A}\bra{A}\hat U^\dag \right).
\end{equation}
All relevant changes need being evaluated between the initial separable state $\rho_{SA}{=}\rho\otimes\rho_A$, and the final state $\rho'_{SA}{=}{\hat U \rho\otimes\ket{A}\bra{A}\hat U^\dag}$. The overall information available in the bipartite state can be decomposed as the sum of three contributions: $I_{\rm tot}=I_S+I_A+I_{S:A}$, where $I_{S:A}$ is the mutual information of the bipartite state~\cite{MikeIke}, and the local information content is $I_S = \ln d - S(\rho)$ for the system, and $I_A = \ln d_A - S(\rho_A) = \ln d_A$ for the ancilla, with  $d\,(d_A)$ the dimension of the Hilbert space of the system (ancilla). Since the evolution is unitary, the total information content of the joint state of the system and the ancilla can not change. Therefore, if we evaluate the new amount of information available after the measurement $I'_{\rm tot}=I'_S+I'_A+I'_{S:A}$, we expect no difference in the total values $I_{\rm tot}=I'_{\rm tot}$, but only a redistribution among the three terms. Here, the final state of the ancilla is given by $\rho'_A= \Tr_S\left( U \rho\otimes\ket{A}\bra{A}U^\dag \right)$. Since the difference in information of the system $\Delta I_S$ equals the variation of its degree of reality up to a sign, we find the relation~\cite{Dieg18}:
\begin{equation}
\Delta I = \Delta I_{S:A}+\Delta I_A = -\Delta{\mathcal I},
\end{equation}
leading to its interpretation as a complementarity relation:
\begin{equation}
\Delta I = -\Delta I_S=  -\Delta{\mathcal R},
\label{complem}
\end{equation}
that rigorously holds only when the coupling is unitary, hence reversible. In this limit, the changes in the degree of reality associated to $O$ are the only source of the variations in the mutual information, and in the marginal information content of the ancilla. 

\begin{figure}[t]
\includegraphics[width=\columnwidth]{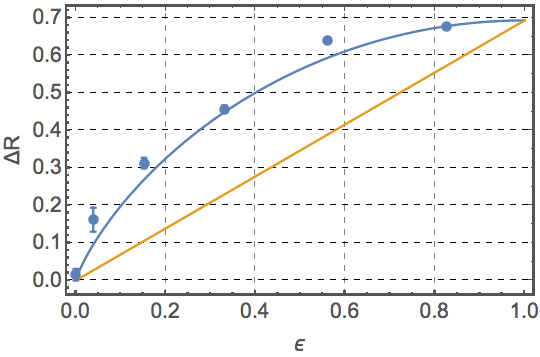}
\caption{Experimental behaviour of $\Delta{\mathcal R}$ as a function of the measurement strength $\epsilon$. The measured values correspond to the points. The blue solid line corresponds to the prediction, and the solid orange line corresponds to the bound \ref{bound}.}
\label{fig1}
\end{figure}

\section{Photonics experiment} We employ the measurement device in~\cite{Pryd04,Pryd05,Manc18} to investigate the experimental behaviour of Eqs.~\eqref{bound} and \eqref{complem} for single qubits. Both system $S$ and ancilla $A$ are encoded in the polarisation of single photons. These interact in an interferometric setup that implements a controlled-phase interaction $\hat U=\ket{0}\bra{0}\otimes{\hat I}+\ket{1}\bra{1}\otimes{\hat Z}$, where $\hat I$ is the 2$\times$2 identity matrix, and $\hat Z$ is the $z$ Pauli operator. This can be used as a weak measurement device of the observable $O=Z$, corresponding to a measurement of the populations of the horizontal $H$ (1) and vertical $V$ (0) components of the system~\cite{Ralp06}. The state of the system is kept fixed in the pure state $\ket{+}=(\ket{0}+\ket{1})/\sqrt{2}$, while the ancilla is initially taken as $\ket{\psi(\theta)}=\cos\theta\ket{0}_A+\sin\theta\ket{1}_A$, and, after the measurement, it is measured in the basis $\{\bra{+}_A,\bra{-}_A\}$. For $\theta=0$ no change is imparted to the ancilla being it an eigenstate of $\hat Z$, hence it will eventually deliver no information on the system; for $\theta=\pi/4$, the ancilla is unaffected, if the system is in $\ket{0}$, and rotated by 180$^\circ$ around the $z$-axis of the Bloch sphere, if the system is in $\ket{1}$. Discriminating between the two possibilities on the ancilla provides full information on the system. In between these two extremes, the state of the ancilla defines the measurement strength as $\epsilon = 1-\cos(2\theta)$~\cite{Ralp06}. In our experiment, we started with a fiducial bipartite state prepared closely to a pure state - we will then assume that the initial entropies are zero with an error comparable to the experimental uncertainties. We then performed full tomography of the bipartite final state, and obtain the relevant quantities in the inequalities \eqref{bound} and \eqref{complem}, following the original suggestion~\cite{Bilo15}.

The results shown in Fig.\ref{fig1} illustrate the measured change in the degree of reality of $Z$ following a weak measurement with tuneable strength. The experimental points have been estimated from the experimentally reconstructed bipartite density matrices $\rho'_{SA}$ after the measurement, tracing out the ancilla. The data follows the predicted behaviour and clearly satisfies the bound Eq.~\eqref{bound}. It is seen how the linear lower limit $\epsilon\,{\mathcal I}(O|\rho)$, for the pure initial state $\rho=\ket{+}\bra{+}$ remains far from the experimental data and from the predictions, except for extremal values of $\epsilon$, where it is most useful.

\begin{figure}[t]
\includegraphics[width=\columnwidth]{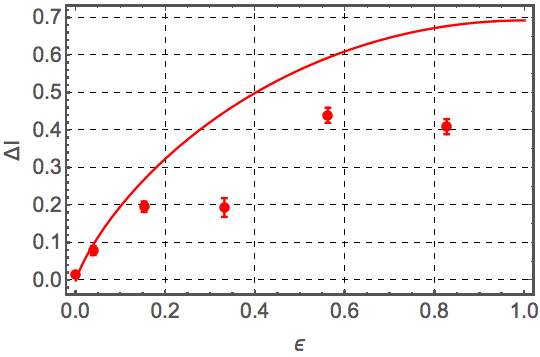}
\caption{Experimental behaviour of $\Delta I$ as a function of the measurement strength $\epsilon$. The measured values correspond to the points. The red solid line corresponds to the prediction \ref{complem}.}
\label{fig2}
\end{figure}

\begin{figure}[t]
\includegraphics[width=\columnwidth]{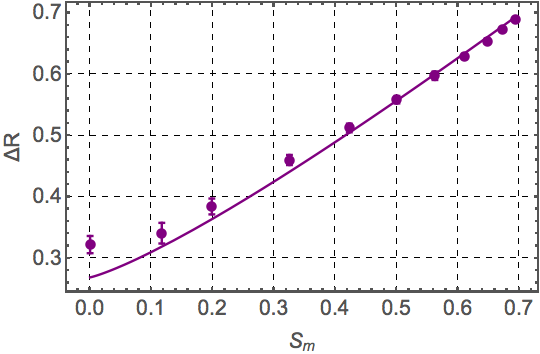}
\caption{Experimental behaviour of $\Delta R$ as a function of the entropy of the initial meter state $S_m$, for $\theta=16^\circ$. The measured values correspond to the points. The purple solid line corresponds to the prediction.}
\end{figure}

We have also evaluated the difference in the information $\Delta I$ from the experimental $\rho'_{SA}$: $\Delta I = S({\mathcal M}^\epsilon_O(\rho))-S(\rho'_{AS})$, due to the fact we start with pure states. The corresponding results are reported in Fig.\ref{fig2}. The experiment shows that the complementarity relation Eq.~\eqref{complem} is more sensitive to external factors, since $\Delta I_S$ saturates at a lower value than expected. This comes from the fact that the coupling between the system and the ancilla photons is not unitary, and the dissipation increases with the measurement strength. Part of the information available is then lost to undetected degrees of freedom of the photon pair acting effectively as an  environment: this, however, still contributes to the emergence of realistic properties, much in the spirit of quantum Darwinism~\cite{Zurek,Corn12,Ciam18}.

The emergence of realism in such open systems can be investigate more systematically by using a mixed meter, {i.e.} a state presenting uncontrolled correlations with the environment. 
This can be mimicked by mixing the counting statistics relative to the state $\ket{\psi(\theta)}$ with that for the orthogonal state $\ket{\psi(\theta+\pi/2)}$~\cite{Manc18} with weights $p$ and $1-p$ respectively: in the latter case, the measurement strength is the same , however, due to the action of $\hat U$, an extra $\hat Z$ rotation is imparted to the signal state after the measurement. This results in increased entropy with respect to that resulting uniquely from the measurement back action~\cite{Manc18,Saga09,Reeb16}. 

In Fig.~3, the data for the variation $\Delta R$ when the system is measured by means of a meter state with initial entropy $S_m{=}H(p)$, where $H(p)$ is the Shannon entropy~\cite{MikeIke}, in the weak measurement regime, $\theta=16^\circ$. As the mixing of the meter increases, the state of the signal starts presenting a more pronounced change in its degree of realism. This is due to the fact that information about the value of $Z$ is present in the meter as well as in the environment to which this was originally correlated: this, clearly, can not be fully retrieved by observing the meter only, but still dictates how realistic properties appear in the system. 

\section{Conclusions}  The matter of establishing when realistic properties emerge in quantum systems can be quantified by using the definition of the degree of irreality ${\mathcal I}(O|\rho)$ in Eq.~\eqref{irreality}, based on Von Neumann entropy. This is equivalent to giving a prominent role to the notion of information: indeed, the degree of irreality so defined is given by the amount of information one needs to describe $\rho$ in full, if the observable $O$ is known. On the other hand, our experiment shows how a metric based on definition  are largely insensitive to the imperfections of the measurement device; however, the impact of the measurement itself can be retrieved by looking at changes in the total information $I_{\rm tot}$ contained in the joint state of system and ancilla. We have also been able to comment on the implications of mixedness in the ancilla: the coupling to an environment makes information on the system available, and this is sufficient for realistic properties to emerge, even if no one can gather it by looking at the ancilla only.

\section*{Acknowledgements} We thank P. Mataloni and F. Somma for discussion. MP acknowledges the DfE-SFI Investigator Programme (grant 15/IA/2864), the H2020 Collaborative Project TEQ (grant 766900), and the Royal Society for financial support.


\begin{thebibliography}{99}
\bibitem{Merm89} N.D. Mermin, Phys. Today 42, 9 (1989).
\bibitem{Bohm52} D. Bohm,  Phys. Rev. 85, 166 (1952).
\bibitem{Cave02}  C. M. Caves, C. A. Fuchs and R. Schack, Phys. Rev. A 65, 022305 (2002).
\bibitem{Ever57} H. Everett, Rev. Mod. Phys. 29, 454 (1957)
\bibitem{GRW} G.C. Ghirardi, A. Rimini, and T. Weber, Phys. Rev. D 34, 470 (1986).
\bibitem{Bass13} A. Bassi, K. Lochan, S. Satin, T. P. Singh, and H. Ulbricht, Rev. Mod. Phys. 85, 471 (2013).


\bibitem{EPR35} A. Einstein, R. Podolski, and N. Rose, Phys. Rev. 47, 777 (1935).
\bibitem{Bell64} J.S. Bell, Physics 1, 195 (1964).
\bibitem{Geno05} M. Genovese, Phys. Rep. 413, 319 (2005).

\bibitem{Puse12} M. F. Pusey, J. Barrett, T. Rudolph, Nature Phys. 8, 475 (2012).
\bibitem{Colb12} R. Colbeck and R. Renner, Phys. Rev. Lett. 108, 150402 (2012).
\bibitem{Lewi12} P. G. Lewis, D. Jennings, J. Barrett, and T. Rudolph, Phys. Rev. Lett. 109, 150404 (2012).
\bibitem{Patr13} M. K. Patra, S. Pironio, and S. Massar, Phys. Rev. Lett. 111, 090402 (2013).
\bibitem{Leif13} M. S. Leifer, Phys. Rev. Lett. 112, 160404 (2014).
\bibitem{Barr14} J. Barrett, E. G. Cavalcanti, R. Lal, and O. J. E. Maroney, Phys. Rev. Lett. 112, 250403 (2014).
\bibitem{Branc14} C. Branciard, Phys. Rev. Lett. 113, 020409 (2014).
\bibitem{Ring15} M. Ringbauer, B. Duffus, C. Branciard, E. G. Cavalcanti, A. G. White, and A. Fedrizzi, Nature Phys. 11, 249 (2015).
\bibitem{Nigg12} D. Nigg, et al., New J. Phys. 18, 013007 (2016).

\bibitem{Bilo15} A.L.O. Bilobran, and R.M. Angelo, Europhys. Lett. 112, 40005 (2015).

\bibitem{Ahar90} Y. Aharonov and L. Vaidman, Phys. Rev. A 41, 11 (1990).
\bibitem{Dieg18} P.R. Dieguez, and R.M. Angelo, Phys. Rev. A 97, 022107 (2018).
\bibitem{Baga16} E Bagan, J.A. Bergou, SS Cottrell, M Hillery, Phys Rev. Lett. 116, 160406 (2016).



\bibitem{Pryd04} G.J. Pryde, et al., Phys. Rev. Lett. 92, 190402 (2004).
\bibitem{Pryd05} G.J. Pryde, et al., Phys. Rev. Lett. 94, 220405 (2005).
\bibitem{Scia06} F. Sciarrino et al., Phys. Rev. Lett. 96, 020408 (2006).
\bibitem{Barb09} M. Barbieri, et al., New. J. Phys. 11, 093012 (2009).

\bibitem{Kim12} Y.-S. Kim, J.-C. Lee, O. Kwon, and Y.-H. Kim, Nat. Phys. 8, 117 (2012).
\bibitem{Roze12} L.A. Rozema et al., Phys. Rev. Lett. 109, 100404 (2012).

\bibitem{Piac16} F. Piacentini, et al., Phys. Rev. Lett. 116, 180401 (2016).




\bibitem{Kim18} Y. Kim et al., Nat Comms 9, 192 (2018)
\bibitem{Manc18} L. Mancino et al., NPJ Quantum Info., 5, 20 (2018).


\bibitem{Vaid96} L. Vaidman, Found. Phys. 26, 895 (1996).


\bibitem{MikeIke}  M.~A. Nielsen and I.~L. Chuang, {\em Quantum Computation and Quantum Information} (2nd ed., Cambridge University Press, 2010).
\bibitem{Ralp06} T. C. Ralph, S. D. Bartlett, J. L. OÕBrien, G. J. Pryde, and H. M. Wiseman, Phys. Rev. A 73, 012113 (2006).

\bibitem{Zurek} W.H. Zurek, Nat. Phys. 5, 181 (2009).
\bibitem{Corn12} M. F. Cornelio, et al., Phys. Rev. Lett. 109, 190402 (2012).
\bibitem{Ciam18} M.A. Ciampini, G. Pinna, P. Mataloni, and M. Paternostro, arXiv preprint 1803.01913 (2018).

\bibitem{Saga09} T. Sagawa, and M. Ueda, Phys. Rev. Lett. 102, 250602 (2009).

\bibitem{Reeb16} K. Abdelkhalek, Y., Nakata, and D. Reeb, arXiv preprint 1609.06981 (2016). 


\end{thebibliography}
\end{document}